\documentclass[12pt]{article} 
\usepackage{epsf}
\usepackage{pazha}
\tightenlines
\def\phflux{phot cm$^{-2}$ s$^{-1}$ keV$^{-1}$}
\def\a{$^{\mbox{\small a}}$}
\def\b{$^{\mbox{\small b}}$}
\def\c{$^{\mbox{\small c}}$}
\def\d{$^{\mbox{\small d}}$}
\def\型{$\pm$}
\def\ergs{$\mbox{erg s}^{-1}$}

\begin{document}

{\it To be published in ``Astronomy Letters'', 2001, v.27, N8} 

\bigskip

\title{\bf X-ray Bursts from the Source A1742-294 in the Galactic-Center
  Region}   

\author{\bf A.A.Lutovinov\affilmark{1}, S.A.Grebenev\affilmark{1}, M.N.Pavlinsky\affilmark{1}, R.A.Sunyaev\affilmark{1,2}}   

\affil{
{\it Space Research Institute, Russian Academy of Sciences, Profsoyuznaya
str., 84/32, Moscow, 117810 Russia}$^1$\\
{\it Max-Planck Institut f{\"u}r Astrophysik, Karl-Schwarzschild-Str. 1,
  85740, Garching, Germany}$^2$}  

\vspace{2mm}
\received{22 Mar 2001}

\sloppypar 
\vspace{2mm}
\noindent

We present results of observations of the X-ray burster A1742-294 near the
Galactic center with the ART-P telescope onboard the Granat observatory. The
shape of its persistent spectra was well described by the model of
bremsstrahlung from optically thin plasma, and it had remained essentially
unchanged over $\sim2.5$ years of observations. We show that the mean
interval between X-ray bursts from the source is several times shorter than
assumed previously, and that the burst profile itself depends on the flux
during the burst. We analyze in detail the strong X-ray burst detected from
this source on October 18, 1990, and trace the evolution of its luminosity
and radiation temperature. \\

\clearpage
 
\section*{INTRODUCTION}

\vskip -5pt 

Because of the high concentration of sources of different nature, the
Galactic-center region is one of the most interesting and most commonly
observed regions in X-rays. Although the first sources in this region were
discovered back in the 1970s (Lewin et al., 1976; Proctor et al., 1978), its
detailed map was first obtained by the EINSTEIN observatory in the soft
X-ray band ($h\nu\le4.5$ keV). These observations revealed more than ten
hitherto unknown X-ray sources and an extended diffuse source (Watson et
al., 1981; Hertz, Grindlay, 1984). Subsequently, soft X-ray observations of
the Galactic center were carried out by the ROSAT (Predehl, Trumper, 1994),
ASCA (Maeda et al., 1996), and BeppoSAX (Sidoli et al., 1999)
observatories. In the hard energy band ($h\nu\ge10$ keV), this region was
fragmentarily observed with the XRT telescope onboard the Spacelab-2
observatory (Skinner et al., 1987) and with instruments of the Spartan-1
mission (Kawai et al., 1988). A more detailed study of the hard X-ray
emission from the Galactic center was carried out with the TTM telescope
onboard the Kvant module (Sunyaev et al., 1991a) and, in particular, with
the ART-P and SIGMA (in gamma-rays) telescopes onboard the Granat
observatory (Pavlinsky et al., 1992a, 1992b, 1994; Sunyaev et al., 1991b).

Persistent X-ray sources of different nature, transients, the millisecond
pulsar SAXJ1808.4-3658, the bursting pulsar GRO J1744-28, and sources in
globular clusters are observed in the Galactic-center region. X-ray
bursters, i.e., neutron stars with weak magnetic fields in low-mass binary
systems on whose surfaces thermonuclear explosions of the accreted matter
occasionally occur, make up a sizeable proportion. The X-ray bursts produced
by explosions are difficult to observe, because the burst source cannot
always be accurately localized and identified with any persistent X-ray
source due to the high concentration of sources in this field. The
ART-P/Granat 2.5-30 keV observations of this sky region are unique not only
because of their long duration (more than $8\times10^5$ s), but also because
of the telescope's technical potentialities. ART-P can image the sky within
its field of view with a high angular resolution and can perform spectral
and timing analyses of any sources in its field of view, irrespective of
their number.

Here, we present our observations of the X-ray burster A1742-294 in the
immediate vicinity ($\sim$1\deg) of the Galactic center. The emphasis is on
the study of peculiarities of bursts from this source.\\

\section*{OBSERVATIONS}

The ART-P X-ray telescope, which is part of the scientific payload of the
Granat international astrophysical observatory, can image a selected region
of the sky by using a coded-aperture technique. It consists of four coaxial,
completely independent modules; each module includes a position-sensitive
detector with a geometric area of $625$~cm$^{2}$ and a coded mask based on
URA sequences. The instrument can image the sky within a
3\fdg4$\times$3\fdg6 field of view (FWHM) with a nominal resolution of
$\sim5$ arcmin (the angular size of the mask element). Because of the
detector's higher spatial resolution ($\sim1.25$ arcmin), the accuracy of
localizing discrete sources is several times higher. The telescope is
sensitive to photons in the energy range 3-60 keV and has an energy
resolution of $\sim22$\% in the 5.9 keV iron line. Readings of the star
tracker, which determines the instantaneous spacecraft orientation with an
accuracy of 1.5 arcmin, are used in imaging and spectral analysis. A more
detailed description of the telescope is given in Sunyaev et al. (1990).

The observations were carried out in the "photon-by-photon" mode, in which,
for each photon, its coordinates on the detector, energy (1024 channels),
and arrival time (the photon arrival time is accurate to within 3.9 ms, and
the dead time is $580\mu$s) were written into the ART-P buffer memory. This
mode allows both timing and spectral analyses of the emission from each
X-ray source within the ART-P field of view to be carried out. Data transfer
to the main memory was made after the temporary buffer was filled (once in
150-200 s) during $\sim$30 s, which resulted in breaks in the information;
therefore, the data were in the form of individual exposures.

The Galactic center was observed by the Granat observatory in series twice a
year, in spring and fall (this was caused by the restrictions imposed on the
satellite in-orbit orientation with respect to the Sun). Five series of
Galactic-center observations were carried out in 2.5 years of the ART-P
operation, with the total observing time being $\sim830000$~s. This allowed
us to analyze in detail the persistent emission from sources in this region,
their spectra and variability on various time scales, to discover several
new X-ray sources, and to detect more than 100 X-ray bursts (Pavlinsky et
al., 1992a, 1992b, 1994; Grebenev et al., 2001).

It should be noted that the first and fourth ART-P modules were used during
the first two series of observations (in the spring and the fall of
1990). The subsequent observations were performed by the third module with a
lower sensitivity in the soft energy band (3-8~keV), which hampered and, in
several cases, prevented the localization and spectral analysis of the X-ray
bursts detected by this module (see Grebenev  et al., (2001) for more
details).\\

\section*{PERSISTENT EMISSION}

The X-ray burster A1742-294 is the brightest persistent X-ray source among
those located near (within $\sim1$\deg) the Galactic center; it is
responsible for $\sim1/3$ of the total flux from this region in the standard
X-ray band (2-20 keV). In this energy band, the source has been repeatedly
observed from various satellites. Based on SIGMA/Granat data, Churazov et
al.~(1995) first showed that A1742-294 also radiates in the hard energy
band ($h\nu\ge35$ keV).

X-ray emission from the source was observed every time it fell within the
ART-P field of view (Pavlinsky et al., 1994). Its 3-20 keV flux was 30-50
mCrab during the 1990 series of observations and slightly decreased to 20-30
mCrab by the spring of 1992. The shape of the burster's persistent spectrum,
which is well described by the model of bremsstrahlung from optically thin
plasma, remained essentially unchanged. The temperature $kT_{\rm br}$ of
this model determined by fitting the spectra of the source in individual
observing sessions was within the range 6.4-10.5 keV. When modeling the
spectra, we took into account their distortion in the soft energy band by
strong interstellar absorption, which is characterized by a hydrogen column
density $N_{\rm H}\simeq6\times10^{22}$ cm$^{-2}$.

As an illustration, Fig. 1 shows the typical persistent spectrum of
A1742-294 measured with the ART-P telescope on September 9, 1990. The dots
indicate the pulse-height spectrum (in counts s$^{-1}$ cm$^{-2}$
keV$^{-1}$), and the solid line represents the corresponding model spectrum
(in \phflux); the temperature that was determined by fitting the spectrum
with the model of bremsstrahlung from optically thin plasma is shown in the
lower left corner.\\

\section*{X-RAY BURSTS}

Lewin et al. (1976) discovered that most bursts detected by them from the
Galactic-center region originated from three sources, A1742-294, A1742-289,
and A1743-28; bursts from the first two sources were observed regularly. Our
analysis of the ART-P data revealed no burst from A1742-289 and A1743-28 (of
course, in those cases where we managed to perform the localization).

\subsection*{Recurrence}

Of 100 X-ray bursts detected over the entire period of ART-P observations of
the Galactic-center region, 26 bursts with a duration of $\sim$15~s were
identified with the burster A1742-294 (see the table). Criteria for the
detection of bursts and their identification with persistent sources can be
found in Grebenev et al. (2001), whence some of the data in this table were
taken. At the same time, note that the burst parameters in the table were
obtained from a spectral analysis of the emission observed during bursts;
i.e., it is a continuation and expansion of the study of bursts initiated by
Grebenev et al. (2001). Two successive bursts at a time were observed during
several sessions, which allowed us to directly measure the characteristic
burst recurrence time $t_{r}$. The measured values of $t_{r}$ lie in the
range 1.5 to 4.2~h with a mean of $\sim2.4$ h, which is several-fold smaller
than the value found by Lewin et al. (1976). The mean energy spectra of the
detected bursts are satisfactorily described by a blackbody model whose
temperature, $kT_{\rm bb}$, changes from burst to burst in the range
$\sim$1-3~keV (see the table) with the mean $kT_{\rm bb}=1.81\pm0.39$~keV.

\begin{table}[t]
\noindent
\centering
{\bf Table 1. }{The X-ray bursts detected by the ART-P telescope from
  A1742-294 during 1990-1992}\\   
\centering
\vspace{1mm}
\small{

\begin{tabular}{r|c|c|c|c|c}
\hline
 Date & $T_{0}$,\a& $kT_{bb}$, & $L_{bb}$,\b     & Flux,\c  & Duration, \\  
      &  UT       & keV        & $10^{37}$ \ergs &  mCrab   & s         \\
\hline
20.03.90  & $18^{h}05^{m}00^{s}$ & 1.64\型0.35 & 2.20\型0.65 & 168\型50& 26\\ 
          & 20 01 29& 1.61\型0.53 & 1.38\型0.69 & 88\型44 & 18\\
24.03.90  & 18 32 42& 2.78\型0.73 & 3.32\型0.92 & 253\型70 & 24\\
          & 21 42 00& 1.87\型0.61 & 2.35\型0.97 & 199\型74 & 17\\
 8.04.90  & 14 06 38& 2.45\型0.34 & 4.26\型0.71 & 281\型47 & 24\\
 9.09.90  & 14 07 48& 1.65\型0.41 & 1.81\型0.92 & 115\型67 & 18\\
          & 15 53 21& 1.89\型0.30 & 3.43\型0.86 & 238\型58 & 17\\
29.09.90  & 13 32 38&  --\d       & --\d        & 372\型90 & 30\\
 5.10.90  & 15 31 44& 2.02\型0.54 & 1.60\型0.69 & 147\型55 & 32\\
          & 19 43 33& 1.43\型0.48 & 3.36\型1.52 & 234\型125 & 21\\
 6.10.90  & 22 10 36& 1.49\型0.23 & 4.91\型1.17 & 374\型89 & 22\\
 9.10.90  & 17 20 52& 2.21\型0.59 & 2.21\型0.75 & 169\型57 & 11\\
          & 19 06 11& 1.60\型0.28 & 2.59\型0.70 & 162\型49 & 15\\
10.10.90  & 14 35 48& 1.97\型0.42 & 2.39\型0.83 &183\型64 & 12\\
          & 16 06 25& 1.55\型0.24 & 2.48\型0.76 & 418\型96 &  9\\
18.10.90  &  9 50 31& 2.15\型0.19 & 7.97\型0.96 & 607\型73 & 17\\
22.02.91  & 14 07 59& 0.97\型0.17 & 8.26\型2.53 & 709\型179 & 13\\
23.02.91  & 22 24 36& 1.98\型0.27 & 7.21\型1.90 & 550\型145 & 16\\
26.02.91  & 11 00 20& 4.79\型2.67 & 2.97\型2.51 & 312\型181 & 15\\
 1.04.91  & 13 39 44& 3.51\型1.39 & 3.30\型1.19 & 252\型91 & 28\\
          & 15 28 57& 1.71\型0.84 & 4.46\型1.77 &343\型136 & 19\\
 8.04.91  & 13 16 25& 1.30\型0.25 & 2.29\型2.00 & 326\型152 & 22\\
15.10.91  & 19 59 08& --\d        & --\d        & --\d      & 14\\
18.10.91  & 11 44 04& --\d        & --\d        & --\d      & 15\\
21.02.92  & 12 52 53& 1.86\型0.43 & 4.51\型1.53 & 283\型115 & 21\\
 2.03.92  &  1 43 22& 2.01\型0.33 &10.06\型3.18 & 963\型239 & 15\\
\hline
\end{tabular}
\vspace{3mm}
 
\begin{tabular}{ll}
\a & The burst peak time (UT).\\
\b & The source's mean 6-20 keV luminosity during the burst at a distance of
8.5 kpc.\\ 
\c & The source's mean 6-20 keV flux during the burst.\\
\d & The source's spectrum cannot be reconstructed by technical reasons.\\
\end{tabular}}
\end{table}

The characteristic burst recurrence time $t_{r}$ can be estimated if the
source's persistent and burst luminosities are known: $t_{r}\simeq
\alpha\Delta t L_b/L_p$, where $L_p$ is the persistent luminosity, $L_b$ is
the mean burst luminosity, $\Delta t$ is the burst duration, and
$\alpha\sim100$ is the ratio of energy release during accretion to that
during thermonuclear helium burning. The derived value, $t_{r}\sim3$ h, is
in satisfactory agreement with direct measurements.

\subsection*{Burst Profiles}

When analyzing the time profiles of the X-ray bursts detected from
A1742-294, we found them to depend on the mean 3-20 keV flux during the
burst: at a mean flux of $\sim100-300$~mCrab, the profile is nearly
triangular, i.e., the rise time of the burst ($\sim5$~s) is close to its
decay time, whereas the burst profile becomes classical, i.e., it is
characterized by a sharp rise ($\sim1-2$~s) followed by a smooth decay
(Grebenev et al., 2001), when the flux increases to $\sim600-1000$~mCrab.
Bursts of the second type occur much more rarely than bursts of the first
type, so two such events in a row have never been observed during a single
session. To analyze this dependence in more detail, we constructed average
profiles for the bursts of each type. Because of the spectral peculiarities
of the third ART-P module, we used only data from the first and fourth
modules. In our subsequent analysis, we rejected bursts that partially
occurred at the beginning or at the end of the exposures. Each of the twelve
remaining bursts was reduced to a dimensionless intensity scale in the 3-20
keV band by its normalization to the difference between the maximum burst
count rate and the persistent count rate determined from the exposure that
preceded the exposure during which the burst was detected. Subsequently, the
dimensionless bursts were added by matching the peaks and averaged. The
results of this averaging for triangular bursts in two energy bands are
shown in Fig.2{\it a}. For comparison, Fig.2{\it b} shows the profile of the
classical burst detected by ART-P on October 18, 1990, in the same energy
bands. This was the only burst of such shape and intensity detected by the
telescope's fourth module. Three more such events were observed by the third
module.

The differences in the burst profiles are clear: whereas the burst in
Fig.2{\it a} has a triangular shape in the soft and hard energy bands
similar to the shape of the burst in Fig.2{\it b} at high energies, the
profile of the latter in the 3-8 keV energy band is strictly classical in
shape: a sharp ($\sim$1 s) rise and a smooth exponential decay. This fact leads
us to conclude that low-energy X-ray emission is responsible for the profile
shape of the X-ray bursts observed from A1742-294. Note also that a
softening of the source's spectrum during the burst clearly shows up in both
figures: the decay time in the hard energy band is considerably shorter than
that in the soft energy band, which is one of the characteristic features of
type I bursts.

\subsection*{The Burst of October 18, 1990} 

On this day, the ART-P telescope detected the strongest and most interesting
burst from A1742-294, during which the peak X-ray 3-20 keV flux was
$\sim1.5$ Crab. Figure 3 shows the source's measured light curves in various
energy bands during this burst. The time profile of the burst suggests that
it belongs to type I bursts, which are caused by thermonuclear explosions on
the neutron-star surface. We see that the peak flux in the hard energy bands
is reached later than that in the soft energy bands; i.e., the burst rise
time in the 12-16 keV energy band ($\sim5$ s) is considerably longer than
the rise time in the 4-8 keV energy band ($\sim1-2$ s). Given that the
e-folding exponential decay time of the burst $t_{exp}$ decreases from 8.7 s
in the 4-8 keV band to 3.9 s in the 8-12 keV band, to 2.3 s in the 12-16 keV
band, and to 1.5 s in the 16-20 keV band, the total burst duration in the
hardest energy band turns out to be several-fold shorter than that in the
softest band (Fig.3). The energy dependence of $t_{exp}$ indicates that the
source's spectrum had softened appreciably by the end of the burst, which is
one of the characteristic features of type I X-ray bursts. Note that no
statistically significant excess of the signal over the background was found
during the burst in the harder energy band ($h\nu\ge20$ keV).

To trace the evolution of the source's parameters during the burst, we
divided it into seven time intervals, reconstructed the photon spectrum for
each of them, and fitted it by a blackbody model. Variations of the source's
temperature and luminosity in the 3-20~keV energy band during the burst
determined in the above model are shown in Fig.4. The source's
burst-averaged spectrum is well described by the radiation law of a
blackbody with a temperature $kT_{bb}\simeq2.15\pm0.19$ keV, a radius
$R_{bb}\simeq 6.4\pm1.4$ km, and a 3-20~keV flux
$F_{bb}\simeq(1.35\pm0.13)\times10^{-8}$ erg cm$^{-2}$ s$^{-1}$, which
corresponds to a luminosity $L_{bb}\simeq1.2\times10^{38}$ \ergs at a
distance of 8.5 kpc. In Fig.5, this spectrum is compared with the
persistent spectrum of A1742-294. The inferred temperature, which
corresponds to the super-Eddington flux, and the small blackbody radius
suggest that the spectrum is distorted by Comptonization, while the dip
clearly seen in the burst profile at high energies apparently points to the
photospheric expansion in the source that took place at the initial burst
stage.\\

\section*{DISCUSSION}

Long-term observations of the Galactic center with instruments of the Granat
observatory have allowed us not only to analyze in detail the timing and
spectral parameters of persistent sources in this region, but also to detect
more than 100 X-ray bursts; we managed to identify 26 of these bursts with
the X-ray burster A1742-294 in the Galactic-center immediate vicinity. Long
continuous observing sessions (more than 10 h) made it possible to directly
measure the burst recurrence time. The derived value, $t_r\sim2.4$ h, proved
to be several-fold shorter than assumed previously.

The X-ray bursts detected by ART-P from this source can be arbitrarily
divided into ordinary (with a mean burst flux of 100-300~mCrab) and strong
(a mean flux of 600-1000~mCrab). The burst profiles differ significantly:
ordinary bursts are triangular (the rise time is comparable to the decay
time), while strong bursts are classical in shape (a sharp rise followed by
a smooth decay).

The mean energy release during the burst did not depend on the source's
pre-burst luminosity and was $E\simeq9\times10^{38}$ ergs for ordinary bursts
and $E\simeq2\times10^{39}$ ergs for strong bursts. To provide such energy
yields, $M\simeq E/\epsilon_{He}\simeq 5.0\times10^{20}$ g and
$1.1\times10^{21}$ g of matter, respectively, must be accreted onto the
neutron-star surface (here, $\epsilon_{He}\simeq0.002c^2$ is the helium
burning efficiency). The rate of accretion onto the neutron star in
quiescence can be represented as $\dot M=L_{\rm X} R_{ns}/GM_{ns}$, where
$L_{\rm X}$, $M_{ns}\simeq1.4 M_{\sun}$, and $R_{ns}\simeq10$ km are the
luminosity, mass, and radius of the neutron star, respectively. Assuming
that the burster's mean luminosity in quiescence is $L_{\rm
X}\simeq8\times10^{36}$ \ergs, we can estimate the characteristic times
[tau] in which the required amount of matter will be accreted onto the
neutron-star surface: $\tau\simeq3.2$ and $7.2$ h for ordinary and strong
bursts, respectively.

The observation of both ordinary (weak) and strong X-ray bursts from
A1742-294 and the fact that the latter are observed much more rarely can be
naturally explained in terms of the above estimates. The distinction in the
burst profiles can be connected with the fact that thermonuclear explosion
responsible for the burst takes place not instantaneously over the entire
neutron-star surface, but only in its region where local conditions for
thermonuclear burning were created, and subsequently spreads over the entire
neutron-star surface. Depending on the amount and density of the accumulated
matter the burning front spreads in the subsonic regime -- deflagration
(weak bursts), or in the supersonic regime -- detonation (strong
bursts). The criteria for realization of these two regimes were investigated
in detail in Fryxell and Woosley (1982). That is the reason why the profiles
of ordinary bursts have a longer rise compared to strong bursts.\\

\section*{ACKNOWLEDGMENTS}

This study was supported by the Russian Foundation for Basic Research
(projects nos. 99-02-18178 and 00-15-99297). 

We wish to thank K.G. Sukhanov, flight director, the staffs of the Lavochkin
Research and Production Center, RNIIKP, and the Deep Space Communications
Center in Evpatoria, the Evpatoria team of the Space Research Institute
(Russian Academy of Sciences), the team of I.D. Tserenin, and B.S. Novikov,
S.V. Blagii, A.N. Bogomolov, V.I. Evgenov, N.G. Khavenson, and
A.V. D'yachkov from the Space Research Institute who operated the Granat
Observatory, provided the scientific planning of the mission, and performed
a preliminary processing of telemetry data. We also wish to thank the team
of M.N. Pavlinsky (Space Research Institute) and the staff of the former
Research and Development Center of the Space Research Institute in Bishkek
who designed and manufactured the ART-P telescope. We wish to thank
V.Astakhov for the help in translating this paper in English.

\pagebreak   

\pagebreak

\begin{figure}[t] 
\hspace{0cm}{
\epsfxsize=150mm
\epsffile[110 375 400 695]{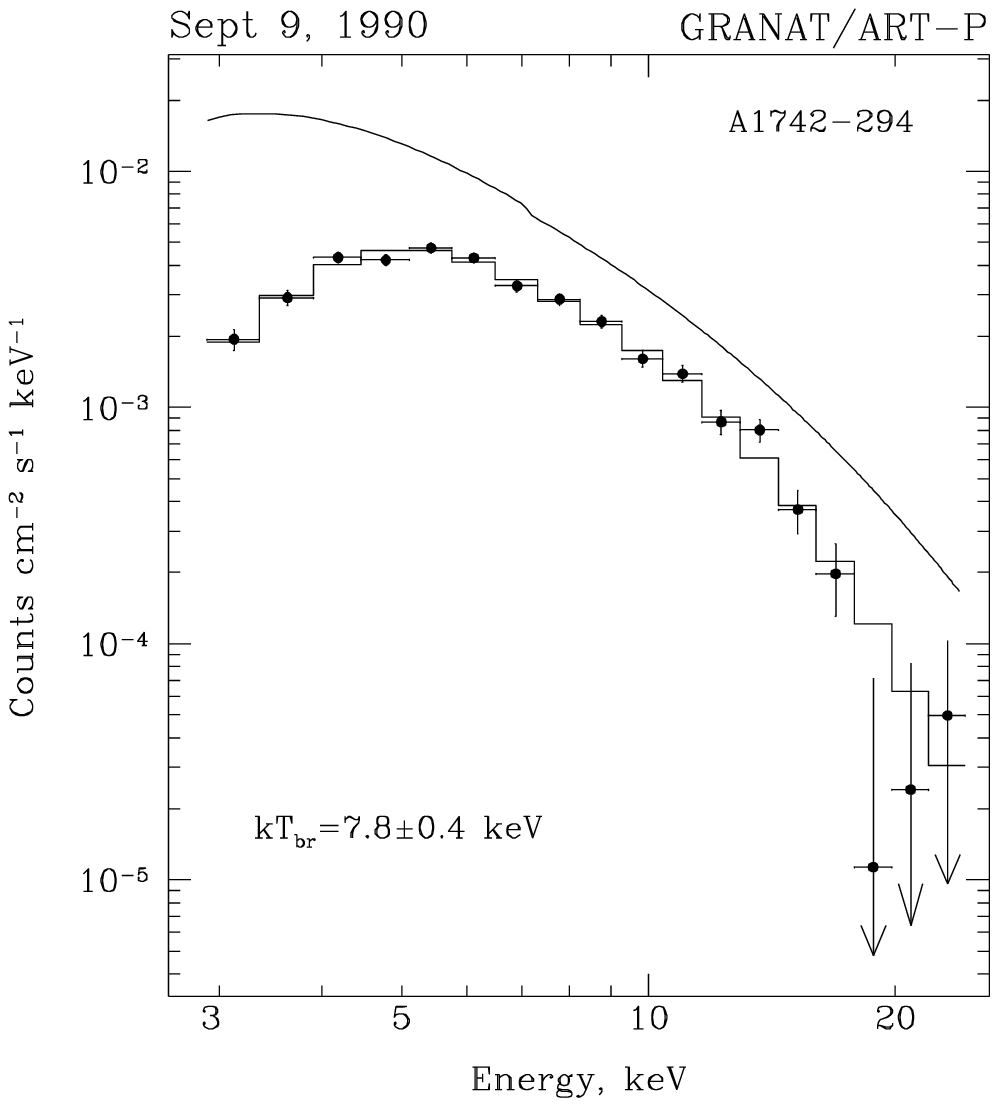}}
\vspace{15pt}
 \caption{\rm The persistent spectrum of A1742-294 obtained with the ART-P
   telescope during the Galactic-center observation on September 9,
   1990. The dots indicate the measured (pulse-height) spectrum, and the
   histogram represents its best fit by the bremsstrahlung model. The model
   (photon) spectrum is indicated by the solid line.} 
\end{figure}

\clearpage

\begin{figure}[t] 
\hspace{-0.5cm}{
\epsfxsize=170mm
\epsffile[35 455 530 685]{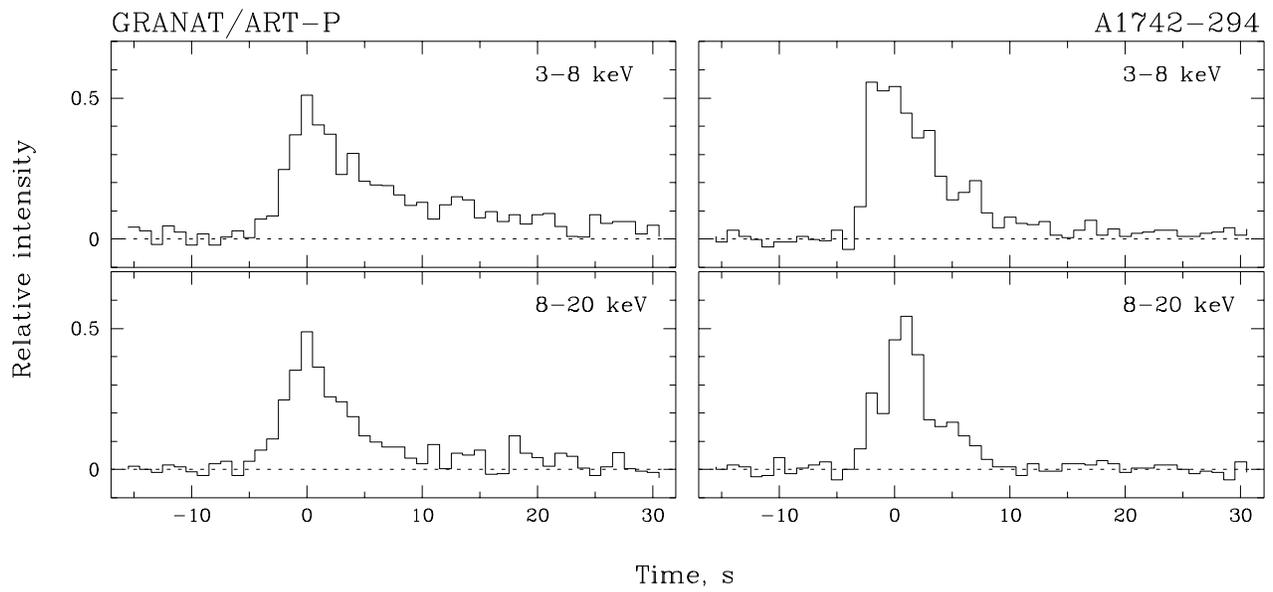}}
\vspace{5pt}
 \caption{\rm The average profiles of bursts with different level of intensity
   observed from A1742-294 in various energy bands.}
\end{figure}

\pagebreak

\begin{figure}[t] 
\hspace{1cm}{
\epsfxsize=120mm
\epsffile[120 285 375 685]{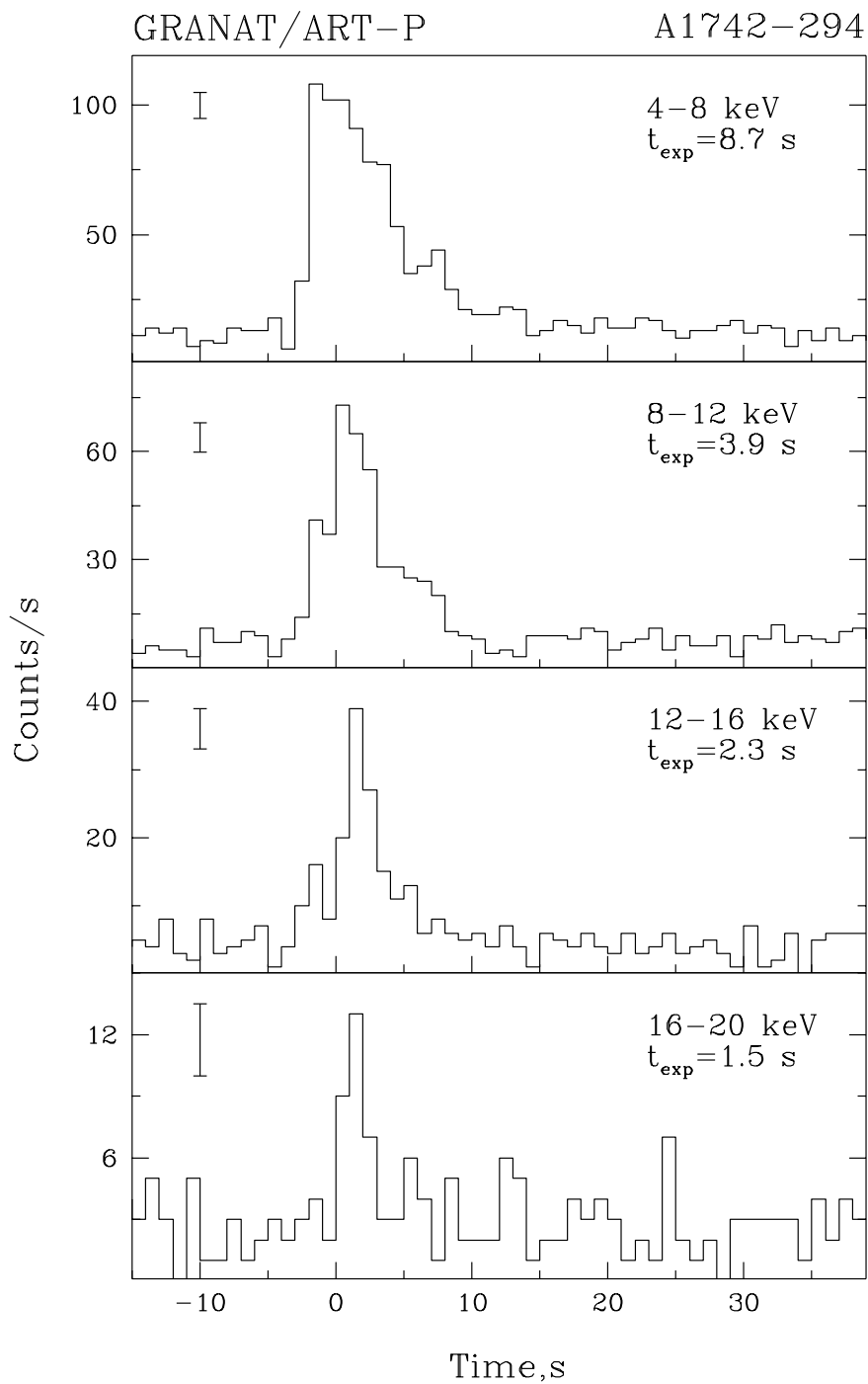}}
\vspace{5pt}
 \caption{\rm The profiles of the X-ray burst detected on October 18, 1990,
   from A1742-294 in various energy bands with a time resolution of 1 s:
   $t_{exp}$ is the source's $e$-folding decay time in each band. The errors 
   correspond to one standard deviation.}
\end{figure}

\pagebreak

\begin{figure}[t] 
\hspace{0cm}{
\epsfxsize=140mm
\epsffile[120 430 375 685]{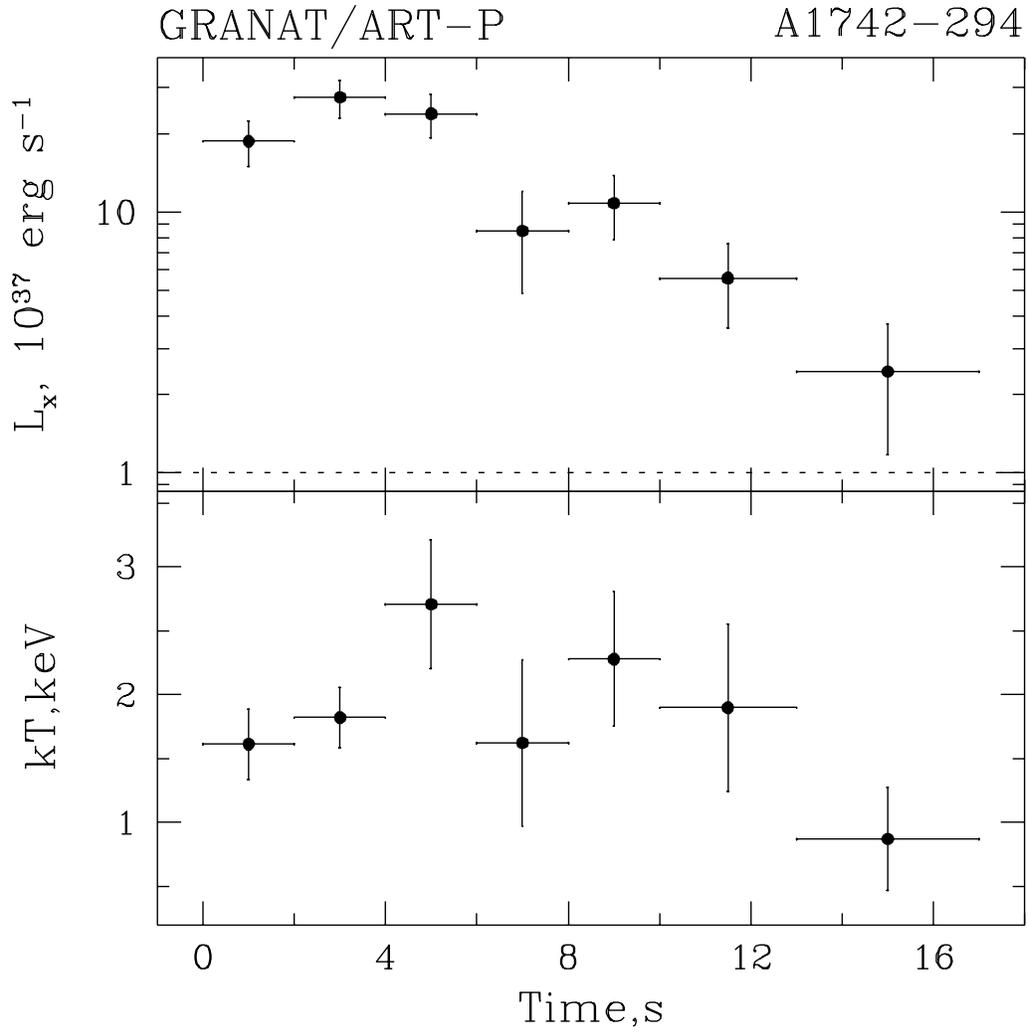}}
\vspace{5pt}
 \caption{\rm Evolution of the source's luminosity and temperature
   determined by fitting the spectra of A1742-294 with a blackbody model
   during the strong burst detected on October 18, 1990. The dotted line 
   indicates the source's persistent luminosity.}
\end{figure}

\clearpage

\begin{figure}[t] 
\vspace{9cm}{
\epsfxsize=170mm
\epsffile[50 280 460 375 685]{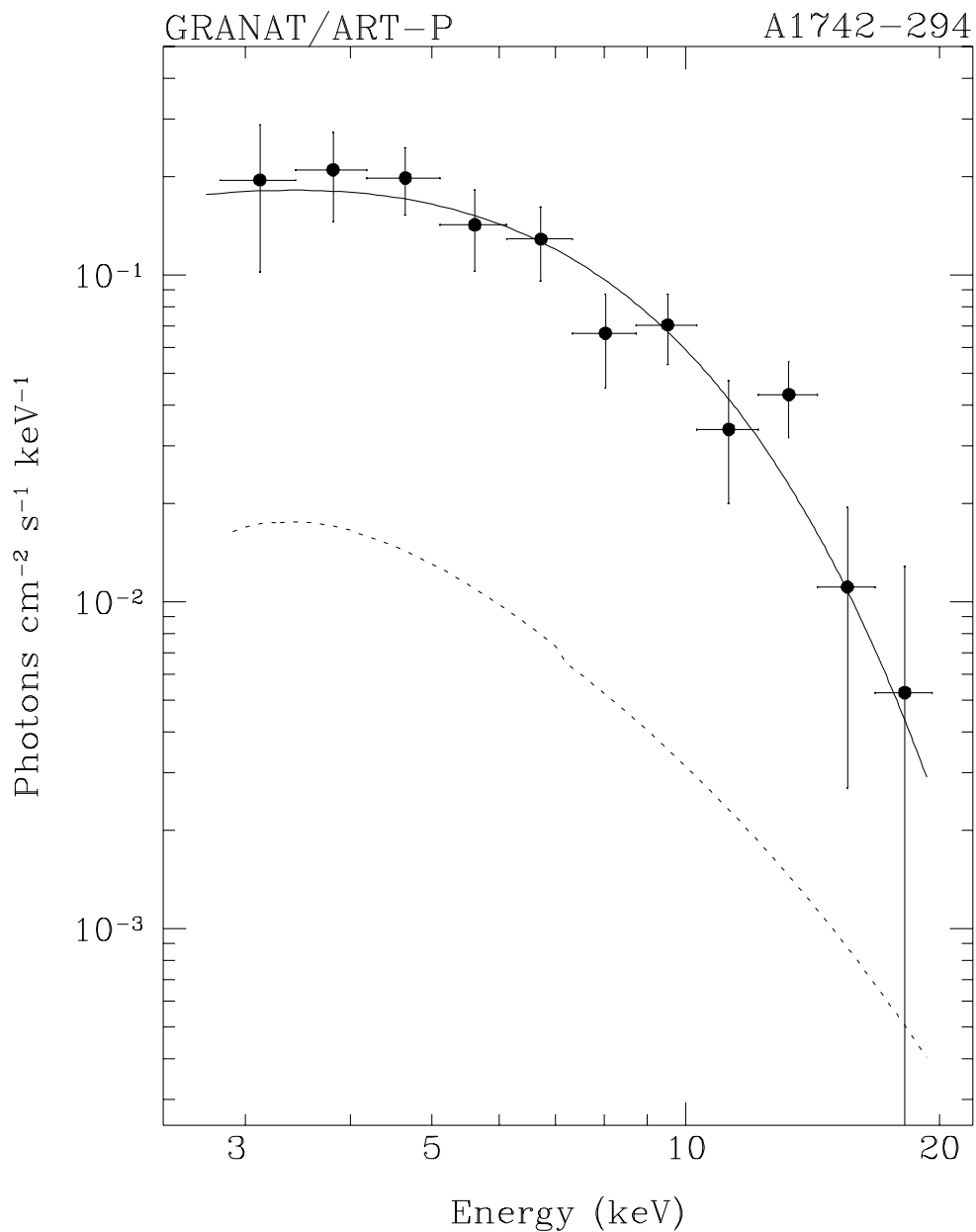}}
\vspace{5pt}
 \caption{\rm The photon spectrum of A1742-294 averaged over the entire October
18, 1990 burst. The solid line represents the model blackbody spectrum that
provides the best fit to the data. For comparison, the dotted line indicates
the source's persistent spectrum.}  
\end{figure}

\end{document}